\documentstyle{l-aa}     

\begin{document}

   \thesaurus{06         
              (02.01.2;02.02.1;  
               08.14.2;
               13.25.5)}
   \title{On the outburst light curves of soft X-Ray transients \\
as response of the accretion disk to mass deposition}


   \author{{\"U}nal Ertan  
   \and M. Ali Alpar} 
   \institute{Physics Department, Middle East Technical University,
     Ankara 06531, Turkey \\
      ertan@newton.physics.metu.edu.tr \\                       
      alpar@newton.physics.metu.edu.tr}

   \maketitle
  \markboth{{\"U}. Ertan \& M. A. Alpar}{{\"U}. Ertan \& M. A. Alpar}
   
   \begin{abstract}
%

 We note that the solution of accretion disk dynamics for an initial 
delta-function mass distribution gives a light curve that fits both 
the rise and the decay pattern of the outburst light curves of black-hole 
soft X-ray transients (BSXTs) until the onset of the first mini 
outburst quite well. The Green's function 
solution of Lynden-Bell $\&$ Pringle~(1974)~is employed for two different 
time-independent viscosity laws to calculate the expected count rates of 
X-ray photons in the Ginga energy bands as a function of time. 
For both models basic characteristics of the outburst light curves of 
two typical sources~GS 2000+25~and ~GS/GRS 1124-68 are reproduced together 
with plausible values of the thin disk parameter $~\alpha~$ and the 
recurrence times. This agreement with the outburst light curves and the 
source properties during quiescence support the idea of mass 
accumulation and the sporadic release of accumulated mass at the outer disk.

\end{abstract}

\def\oneskip{\vskip\baselineskip}
\def\exo{{\sl EXOSAT }}
\def\ein{{\sl EINSTEIN }}
\def\uhu{{\sl UHURU }}
\def\ro{{\sl ROSAT }}
\def\ergsec{\hbox{erg s$^{-1}$ }}
\def\ergcm{\hbox{erg cm$^{-2}$ s$^{-1}$ }}
\def\anr{\it Ann. Rev. Astr. Astrophys., \rm}
\def\nat{\it Nature, \rm}
\def\mn{\it Mon. Not. R. astr. Soc. \rm}
\def\apj{\it Astrophys. J. \rm}
\def\apl{\it Astrophys. J. (Letters) \rm}
\def\aps{\it Astrophys. J. Suppl. Ser.) \rm}
\def\anr{\it Ann. Rev. Astr. Astrophys. \rm}
\def\aaa{\it Astron. Astrophys. \rm}
\def\aj{\it Astron. J. \rm}
\def\aa{\it Acta Astr. \rm}
\def\pasp{\it Publ. Astr. Soc. Pac. \rm}
\def\ssr{\it Space Sci. Rev. \rm}
\def\prl{\it Phys. Rev. Lett. \rm}
\def\paj{\it Publ. Astron. Soc. Japan \rm}
\def\la{\raise.5ex\hbox{$<$}\kern-.8em\lower 1mm\hbox{$\sim$}}
\def\ma{\raise.5ex\hbox{$>$}\kern-.8em\lower 1mm\hbox{$\sim$}}
\def\ea{\it et al. \rm}
\def\am{$^{\prime}$\ }
\def\as{$^{\prime\prime}$\ }
\def\eg{{\sl EGRET }}
\def\be{\begin{equation}}
\def\ee{\end{equation}}
\def\ba{\begin{eqnarray}}
\def\ea{\end{eqnarray}}
\def\d{\partial}
\def\R{\right}
\def\L{\left}
\def\D{\mit\Delta}
\def\S{\Sigma}
\def\bc{\begin{center}}
\def\ec{\end{center}}

\def\xr#1{\parindent=0.0cm\hangindent=1cm\hangafter=1\indent#1\par}

\section{Introduction}

~~~~About two thirds of the known LMXBs are persistent and the 
remaining one third are transient sources (van Paradijs 1995). The 
transient sources exhibit a soft X-ray spectrum during outbursts (soft 
X-ray transients $-$SXTs). An optical outburst also occurs together with 
the X-ray outburst. During 
an outburst the X-ray and the optical properties of black-hole soft 
X-ray transients (BSXTs) and neutron star soft X-ray transients (NSXTs) 
are very similar to each other and also to those of persistent sources. 
X-ray luminosities  of~ SXTs~ increase from below 
$~10^{33}~$erg~s$^{-1}$ to $~\sim10^{37}$-$10^{38}$ erg~s$^{-1}$~ during 
an outburst. The time scales for decline range from 
several tens of days to more than one hundred days. 
Although some sources show rather monotonic decline in 
many cases there is more complex behavior. Some sources remain persistent 
for more than one year. Most of the light curves show a steepening around 
$10^{36}~$erg~s$^{-1}$ corresponding to a mass accretion rate 
$\dot{M}_{x}~\sim10^{16}~$g~s$^{-1}$ inferred from the X-ray luminosity. 
Later they turn back to 
their quiescent states with $L_{x}\sim10^{31}-10^{33}~$erg~s$^{-1}$ and 
$\dot{M}_{x}\sim10^{11}-10^{13}$ g~s$^{-1}$~ (Tanaka $~\&~$ 
Shibazaki~1996~ and references therein).

There are nine known X-ray binaries with strong black-hole candidate 
primaries (van Paradijs 1995). Three of these sources are non-transient 
high mass systems and the remaining six are transient LMXBs~(Table 1). 
Four BSXTs in this group, A0620-00, GS 2000+25,~GS/GRS 1124-68~ 
and GRO J0422+32~ show striking similarities. 
Their outbursts have fast rise times 
around a few days. The decay phases can be fit with exponentials 
with time 
constants around one month. During decline all four sources exhibit 
secondary (mini) outbursts. A0620-00, GS/GRS~1124-68 show also tertiary 
outbursts. The tertiary outburst is absent in the light curve 
of ~GRO J0422+32, but it may also be present in GS~2000+25. 
Secondary maxima were detected around 80 days after the onset of the main
outburst of GS~2000+25 and GS/GRS~1124-68. A0620-00 and GRO J0422+32
exhibited the secondary maxima 50 days and
125 days after the onset respectively.
For both GS~0620-00 and GS/GRO~1124-68 the tertiary maxima were
observed about 200 days after the onset of the main outburst (Tanaka $\&$
Shibazaki~1996). The basic features of the mini outbursts are
the sharp increase in luminosity by a  factor of ~$\sim 1-3$~ and decay
patterns which mimic the decay after the first outburst. 
A0620-00~ is a recurrent transient with a recurrence time of ~58~ years 
~(Tsunemi et al 1989). The outbursts of the other sources were 
detected only once putting a lower limit to their recurrence times of 
around a few ten years~ (see Table 1). These similarities allow the 
working hypothesis that all of these sources run with a similar 
mechanism.

Since the conditions for outburst build up in the 
quiescent state it is important to treat the observations made 
during quiescence. According to the standard disk model 
(Shakura $\&$ Sunyaev 1973) most of 
the X-rays ($>70\%$) come from the inner parts of the disk extending to a 
radius which is around ten times the radius of 
the last stable orbit, $~R_{0}=3R_{g}~$ 
$=3(2GM_{\mbox{x}}/c^{2})$. During the quiescent state it is possible to 
fit a black-body curve roughly to 
the observed X-ray spectrum. However, inner 
disk temperatures of$~\sim 0.2-0.3~$keV obtained from these fits 
give very  small values ~($1-10~$km$^{2}$) for the X-ray emitting area of 
the disk, while realistic areas imply temperatures that are orders of 
magnitude smaller than the soft X-ray temperatures that fit the spectra. 
A disk becomes optically thin below a certain accretion rate depending on 
the viscosity parameter $\alpha$ (Shakura $\&$ Sunyaev 1973). An 
optically thin or gray disk may be one way of explaining this inner disk 
problem. 

The mass accretion rates obtained from the 
optical observations during quiescence 
($\sim10^{15}-10^{16}~$g~s$^{-1}$) are 
orders of magnitude larger than those ($\sim10^{11}-10^{13}~$g~s$^{-1}$)  
obtained from the X-ray luminosities (see, e.g., McClintock et al 1995, 
for these questions in  A0620-00). 
 
An advection dominated inner disk has been proposed to account for the 
properties of BSXTs. According to this model most of 
the energy released in the inner disk is advected into 
the compact star instead of being radiated from the 
disk (Narayan et al 1996). Regarding the difference between mass
accretion rates obtained from optical and X-ray luminosities during the
quiescent states, the NSXTs are similar to BSXTs. Although this model 
could explain the observed 
spectra and also the low X-ray luminosities of BSXTs in quiescence it is not 
able to 
explain the properties of NSXTs because of the existence of the solid 
surfaces of the neutron stars: Whatever the disk structure is matter 
finally reaches the neutron star surface which should produce a distinct  
black-body like emission. 
 
Another possibility is mass accumulation in the outer regions of the disk. 
This accumulated matter may be the source of the outburst. For both 
BSXTs and NSXTs, the mass accretion rate inferred from optical 
observations may indicate the mass accretion rate arriving to accumulate 
in the outer disk, while the much lower $\dot{M}_{\mbox{x}}$ inferred from 
the X-ray luminosity may reflect the trickle of mass 
that proceeds through a mass-starved inner disk 
characteristic of the quiescent state. This qualitative picture would 
persist until the mass accumulated in the outer disk reaches the  
level critical for an outburst.  

Models for SXTs must address the characteristic 
time scales: (1) Fast rise of the light curves 
around a few days; (2) The subsequent decay with a time scale of the 
order of one month; (3) Recurrence times of the order 
of a few ten years. Further, as viscosity is involved in determining the 
decay, values of the viscosity parameter must be plausible. For 
comparison $\alpha\sim 0.1-1$ for dwarf novae  and related systems for the 
reaction time of the disk to be $10^{5}-10^{6}~$s (Bath $\&$ Pringle 1981).

There are mainly two types of models for SXTs: the disk instability 
models (DIM) ~(e.g. Meyer $\&$ Meyer-Hofmeister 1981; Mineshige $\&$ 
Wheeler 1989; Cannizzo 1992; Cannizzo et al 1995~(CCL)) and the  
mass transfer instability models (MTI) (e.g. ~Hameury et al 1986; 1987; 
1988; 1990).

The basic idea for MTI is that the accretion from the secondary star is 
unstable for a range of accretion rates roughly from  
$\sim 10^{12}-10^{15}~$g~s$^{-1}$~to~$\sim~10^{16}-10^{17}~$g~s$^{-1}$. 
The secondary star expands under the influence of hard X-rays 
($E > 7~$keV) from the primary. The accretion rate exceeds the 
upper limit of the lower stable range and jumps to an accretion rate
greater than the minimum of the higher stable range.  
The subsequent  burst of mass flow produces 
the outburst. The disk becomes thicker and thicker, finally shielding the 
region around the $L_{1}$ point. When the X-ray illumination stops 
the companion shrinks and the system returns to its quiescent state. 
MTI models produce the basic characteristics of the light curves and the 
recurrence times. The problem with MTI is that, according to the 
model, the hard X-ray flux with $E>$7~keV~ must exceed 
$\sim 2.5 \times 10^{34}~$$ (M_{c} / M_\odot)^{2}~$erg~s$^{-1}$ 
in the quiescent state (Hameury et al. 1986) while there is no 
data showing the existence of photons in this energy 
range with such high luminosities before the outbursts 
(Tanaka $\&$ Shibazaki~1996). 

DIMs are highly accepted today, although they do not 
produce a completely self consistent picture explaining all features of 
SXTs. Initially DIMs were successful in explaining basic 
characteristics of dwarf novae (Cannizzo 1994). The model was extended 
to explain the behavior of SXTs which 
are similar to dwarf novae in some cases. The disk instability follows a 
limit cycle mechanism based on an "S" shaped surface 
density $\Sigma$ versus $\dot M$ curve. Upper and 
lower branches of the "S" correspond to hot and cool states respectively. 
The middle branch represents an unstable regime. 
When ~$\dot{M}>~\dot{M}_{min}~$ the disk follows the upper branch 
switching to the lower branch when ~$\dot{M}<~\dot{M}_{max}$.  
Each radius in the disk has its own "S" curve. If the mass 
transfer satisfies the condition $~\dot{M}_{max}$$(R_{inner})<$ 
$\dot{M}_{T} <$ $\dot{M}_{min}$$(R_{outer})~$ then the limit cycle 
mechanism operates through the range between $R_{inner}$ and 
$R_{outer}$. If this range of radii covers the entire disk then the whole 
disk can jump from one stable branch to the other.
Accumulation of matter at some radius R may cause mass per unit area 
$\Sigma(R)$ to exceed $\Sigma_{max}(R)$. 
Viscous dissipation increases suddenly. Waves of 
surface density propagate to both smaller and larger radii. As a 
result, the surface density at each radius in the disk exceeds the maximum 
critical value of the lower branch, and 
the disk jumps to the hot branch. At the end of this 
process the disk finds itself in a high viscosity state. 
Unlike the quiescent state now the viscous time scale becomes small 
and matter flows on to the central object. Because of the matter flow the 
densities in the outer regions decrease to below the critical densities. 
This causes a cooling front to propagate throughout the disk decreasing 
the local surface density at each radius to below the local critical 
values. Propagation of the cooling front means the decay of the outburst
(e.g., Cannizzo 1992; Lasota et al 1996).

The main difficulty of this model is that the $\alpha$ values needed in 
order to produce recurrence times of around a few ten years are not able  
to produce the observed amplitude and duration of the outbursts or vice 
versa~(for a discussion of the problems see, e.g.  Lasota 1996).

In this work  BSXTs will~be studied with  a different approach
concentrating on a simple explanation in terms of a disk dynamics model 
for both rise and the decay pattern of the 
light curves until the onset of the first mini outburst. We explore the 
behavior of the 
disk following a sudden release of mass in the outer radius in terms of
the disk dynamics model of Lynden-Bell $\&$ Pringle (1974) (LP). 
The observations 
of GS 2000+25 and GS/GRS 1124-68 will be used to illustrate the model. 

LP made a study for the evolution of viscous disks in general. In
particular the solution with no central flux is of interest,
corresponding to a disk with a black-hole at its center. The Green's 
function solution of LP take initial density 
distributions  in the form of delta functions in radius. This is an 
idealization of an initial mass enhancement in a ring. 
Convolution of the  elementary solution of this Green's function
model with any initial mass distribution will give the general solution.  
Initial mass release at a thin ring at the outer radius (or at a 
specific radius in the disk) already corresponds to 
a solution with a single delta function initial mass distribution. Thus  
the Green's function solution of LP is taken here as the physical
solution for the BSXT outbursts. The motivation leading us to this approach
is the similarity between the observed X-ray light curves of the
outbursts and the Green's function luminosity calculated by LP. This
hypothesis means that a sudden dumping of mass in the outer disk leads to 
the  observed outburst. The present work applies the model to the 
data from ~GS 2000+25~and~GS/GRS 1124-68.

The part of the study by LP related to our application is 
summarized in the Appendix. In Section 2 data from GS~2000+25 and 
GS/GRS~1124-68 are examined with the model.
The aim is to calculate the photon flux from the
disk in the observed X-ray band of Ginga ($1-20~$keV) as a function of time 
using the LP model, and compare this with observed 
photon flux data taken with the Ginga satellite ( provided by 
S.Kitamoto, private communication). A good fit to the data until the 
first mini outburst is obtained (Figs. 1$-$4). From this fit the
characteristics of the  outburst, the viscosity parameter $\alpha$, the
total mass released and the recurrence times will be obtained for both 
sources. Section 3 summarizes the conclusions, discusses the scope of the 
model and the problems, and tests for future applications, including the 
secondary and tertiary 
maxima commonly seen in the decay phases of the outbursts of BSXTs.
 
\begin{table*}
\caption[]{BSXTs with $~M_{\mbox{x}} > 3M_{\odot}$~
(Ref:~Tanaka $\&$ Shibazaki 1996 and references therein)}
\begin{center}
\begin{tabular}{|llllll|}
\hline
~Name~~~&~~BH~mass~~~&~~Outburst
year~~~&~~$L_{\mbox{ave}}$~~&~$\langle\dot{M}\rangle$~& \\
~&$~(M_{\odot})$&&($10^{35}~$erg/s)&$(10^{15}$~g/s)&\\
\hline
J0422+32 &$ >3.2$ &1992&0.8 &0.2 &  \\
0620-00~ & $>7.3$ &1917,1975 &2 &2 &  \\
1124-68~ & $\sim~6$ &1991&7 &7&  \\
J1655-40 & 4-5 &1994&? &? &  \\
2000+25~ & 6-13.9 &1988 &3 &3 &  \\
2023+33~ & 8-15.5 &1956,1979~? &2~? &2~? &  \\
&&1989 &0.6&0.6&  \\
\noalign{\smallskip}
\hline
\end{tabular}
\end{center}
\end{table*}

\section{Models For the X-Ray Light Curves of Black-hole Soft X-Ray 
Transients}

LP present a luminosity versus dimensionless time,~$t_{*}$, curve for
the delta function initial condition. This curve is similar to the 
observed light curves of BSXTs. 
The similarity is attractive because a mass
accumulation at the outer disk may also be represented by a delta 
function initial mass distribution. If a disk instability causes a 
transition to a high viscosity  
state so that the viscous time scale becomes short compared to that in 
the 
preoutburst state then a sudden release of mass would occur. For such a
situation the Green's function solution with  
a delta function initial mass distribution would approximately describe 
the subsequent evolution of the disk.

To test this idea we consider a delta function initial mass distribution
located at a radius $R_{1}$, $~\Sigma(R_{1},R,0)= \delta m~(2\pi R)^{-1}$ 
$\delta(R-R_{1})~$, and use the Green's function in Eq.($A.18$) 
as the solution of our problem. $~\delta m~$ is the mass which is initially 
injected at the outer radius ~$R_{1}$. 
The solution is equivalent to Eq.($A.18$) except for a factor
~$\delta m~$ in front. 

We try two viscosity models and compare the results to the observations.
In the first model we choose the kinematic viscosity $\nu$ to be 
constant for simplicity. In the second model we use the form of the viscosity
parameter~ $\alpha= \alpha_{0}(z_{0}/R)^{n}$~where $z_{0}$ is the half  
thickness of the disk. This form is commonly used for the disk instability 
models. CCL succeeded to produce exponential decays of the light 
curves of BSXTs using this form with $n=1.5$. For the LP model 
Eq.($A.10$) is valid if $\nu$ is time independent, and  either independent 
of $R$ or a power law function of $R$. 
On the other hand the general prescription of the kinematic viscosity is 
$\nu=(2/3)(P/\rho)(\alpha/\Omega)$. Since $(P/\rho)~\propto~T~$,~~$\nu$ 
becomes independent of $T(R,t)$ only when ~$\alpha~\propto~T^{-1}$. This 
choice corresponds to $~n=-2~$and leads ~$\nu$~to be a function of 
the specific angular momentum $h$ as we will 
discuss in the second model. 
For both models the photon flux is calculated as a function of time, as 
the injected mass at $R_{1}$ spreads through the disk. The choice of 
~$\alpha~\propto~T^{-1}$ may not be physical. It is implemented here in 
order to apply the LP model. The success of the model may indicate that a 
real disk is effectively represented by the LP model with the artificial 
choices made here.  Although a 
temperature and time independent viscosity is unrealistic, the success of 
the simple analytical LP solution in fitting the light curve motivates us 
to take this prescription as a useful effective representation. We shall 
discuss the viscosity parameter $\alpha$, the accumulated 
mass $\delta m$~ and the recurrence times using the two viscosity models. 

The models discussed below are summarized in Tables (2$-$4)

\newpage

{\bf MODEL I : Constant ~$\nu$}

We first investigate a model with $\nu$ constant in time and uniform 
throughout the disk. We shall obtain limits on $\nu$ by requiring that the 
(variable) $\alpha$ parameter is less than one. 
If the initial mass distribution is given by Eq.($A.16$) multiplied by 
$\delta m$ the general time dependent solution for the couple is 
Eq.($A.18$) multiplied by 
$\delta m$ (see Appendix)
\be 
g=\delta m~ 2l\kappa^{-2}~\frac{(x_{1} x)^{l}}{x_{1}^{2}
t_{*}}~exp-\left[\frac{(x_{1}-x)^{2}}{2t_{*}   
x_{1}^2}\right]~F_{l}\left(\frac{x/x_{1}}{t_{*}}\right) \label{u2}
\ee
where $l$ is constant, 
$x=h^{1/2l}=(\Omega R^{2})^{1/2l}= (GMR)^{1/4l}$,~ $t_{*}=$ 
$2\kappa^{-2} t~/~ x_{1}^{2}$, $~F_{l}(z)= e^{-z}~I_{l}(z)$, 
$I_{l}(z)$~is the Bessel function of imaginary argument, and 
\be
\kappa^{-2}= 3~(GM)^{2}~\nu \label{u3}~~. 
\ee
The dimensionless time is
\be
t_{*}= \frac{6\nu t}{R_1^{2}} \label{u4}~~.
\ee
Substituting 
\be
\frac{dh}{d\Omega}= \frac{1}{2}~(GM)^{1/2}~R^{-1/2} \label{u5}~~
\ee
in Eq.($A.10$) we have
\be\frac{3}{4}~(GM)^{2} \nu h^{-2}= 4l^{2} \kappa^{-2} h^{2-(1/l)} 
\label{u6}~~.
\ee
Then $l=1/4$ for constant~$\nu$. For~ $0~<~\frac{(x/x_{1})}{t_{*}}~\ll~1$
\be
I_{l}\left[\frac{(x/x_{1})}{t_{*}}\right]\simeq \frac{1}{\Gamma 
(l+1)}~\left[\frac{(x/x_{1})}{2 t_{*}}\right]^{l} \label{u7}~~.
\ee 
Substituting $l= 1/4$, $\kappa^{-2}$ and $I_{l}$~ in Eq.($A.18$) 
we obtain the energy dissipation per unit area
\ba
D&=&~\frac{1}{2\pi R}~g~\frac{\d\Omega}{\d R}~ \nonumber \\
&\propto&~
R^{-3}~t_{*}^{-5/4}~exp\left\{\frac{-[1+(R/R_{1})^2]}{2~t_{*}}\right\} 
\label{u9}. 
\ea
This contains a sharp rise in the exponential factor, and decays 
approximately as a power law at late times. 
The effective temperature of the disk is 
\ba
T&=&~\left(\frac{D}{2~\sigma}\right)^{1/4}~ \nonumber \\
&=&~c~R^{-3/4}~t_{*}^{-5/16}~exp\left\{\frac{-[1+(R/R_{1})^2]}
{8~t_{*}}\right\}\label{u10}
\ea
where $\sigma$~is Stefan-Boltzmann constant and 
$c\simeq~3.9\times10^{7}~(\delta m~\nu~R_{1}^{2})^{1/4}~$(cgs).
To find the total luminosity radiated in the disk as a function of  
time, D must be integrated over the surface of the disk, that is,
\ba
L(t_{*})=\int_{R_{0}}^{R_{1}}~D~2\pi R~dr \nonumber~~.
\ea

For the X-ray luminosity $L_{\mbox{x}}$ to be determined in the 
observational energy band from ~1~to~20~keV, the temperature 
in Eq.(\ref{u10}) is substituted in the Planck function
\be 
{\it F}_{\epsilon}= \frac{2\pi}{h^{2} 
c^{2}}~\frac{\epsilon^{3}}{e^{\epsilon/kT}-1}  \label{u11}
\ee
and a numerical integration must be performed over the X-ray band 
throughout the disk. Since
\ba
{\it F}_{\epsilon}= {\it F}_{\epsilon}~ [~\epsilon,~T(t_{*}, R)] \nonumber~~,
\ea 
the luminosity is found for each annular section of the disk with 
thickness dR at a distance R 
from the center and we find the total X-ray luminosity of the disk by 
integrating. 

The observations report the X-ray photon flux in the 1$-$20 keV band 
rather than the X-ray luminosity.  
We use the photon flux 
\be
{\it N}_{\epsilon}= \frac{2\pi}{h^{2}
c^{2}}~\frac{\epsilon^{2}}{e^{\epsilon/kT}-1}  \label{u12}~~
\ee
for numerical integration to calculate the total photon flux expected 
from the disk as a function of time, $t_{*}$. 
From Eq.(\ref{u10}) the temperature of the inner disk is
\be
T_{0}= 
c~R_{0}^{-3/4}~t_{*}^{-5/16}~exp\left\{\frac{-1}{8~t_{*}} 
\right\} \label{u15}
\ee
since $R_{0}/R_{1} \ll 1$. We designate the dimensionless time $t_{*}$ 
corresponding to the maximum of the light curve by $t_{*}$(max):
\ba
\left[\frac{\d T_{0}}{\d t_{*}}\right]_{t_{*}=t_{*}\mbox{(max)}}~=~ 0 
\nonumber \ea
and we find $t_{*}$(max)$=0.4$. If we set 
$T_{0}=T_{0}$(max)~when~$t_{*}=t_{*}$(max)~then
\be
c\simeq~1.027~R_{0}^{3/4}~T_{0}\mbox{(max)} \label{u16}~~.
\ee
Taking the last stable orbit as the inner radius, $R_{0}= 3R_{g}$~ where 
$R_{g}= (2GM/c^{2})$, Eq.(\ref{u10})~becomes
\ba
T= 1.027~R_{0}^{3/4} T_{0}\mbox{(max)}~R^{-3/4}t_{*}^{-5/16}~\nonumber \\
  \times exp\left\{\frac{-[1+(R/R_{1})^2]}{8~t_{*}}\right\}
\label{u17}~~.
\ea

$T$ is substituted in Eq.(\ref{u12}) and numerical integration is 
performed over the observed X-ray band~$(1-20~$keV)~throughout 
the disk.

The total X-ray photon flux$~n~$received by the detector is 
\be
n=\frac{cos~i}{2\pi d^{2}}~N \label{u18}~~
\ee
where $N$ is the total X-ray photon flux radiated by the visible 
surface of the disk, $d$ is the distance of the source and $i$ is the 
inclination angle between the normal of the disk and the line of sight 
of the observer. 

~~~~~~~

{\bf i.~GS 2000+25}

The distance of ~GS 2000+25~ is around 2 kpc. The black-hole mass is 
$~M_{\mbox{x}}=(5.44\pm 0.15)~sin^{-3}~i~M_{\odot}$  (Harlaftis, Horne 
~$\&$~Filippenko 1996). Taking the inclination angle 
$~47^{\circ}<~i~<75^{\circ}$~, the mass range 
is $~6.04~<~(M_{\mbox{x}} / M_{\odot})~<~13.9~$ (Harlaftis, Horne 
~$\&$~Filippenko, 1996)
   
Setting the inner radius at the last stable orbit,~$R_{0}= 3R_{g}$~,  
the outer radius of the disk ~$R_{1}$~may be 
chosen $\sim10^{3}-10^{4}~$times the inner radius. 
Numerical integration for photon flux is not sensitive to the outer 
radius~$R_{1}$~as long as ~$R_{1} \gg R_{0}$,  
since  most of the X-ray radiation comes from the inner part of the 
disk. The maxima of the model light curve and the observed light curve 
can be matched by adjusting the single parameter ~$T_{0}$(max). This must 
be done for different masses tracing the possible $M_{\mbox{x}}$ range.
The values of ~$T_{0}$(max)~ corresponding to different masses 
and different inner radii are given in Table(2).

Writing the dimensionless time as $~t_{*}= b~t~$
we obtain the value of the scale factor $b$ that gives the best fit 
(minimum $\chi^{2}$) between the model 
and the observed photon count light curves 
for a grid of mass values in the allowed range. 
We do this after matching  the maxima of the 
theoretical and the observed light curves, since there is not enough 
data during the initial increasing phase of the photon flux. The 
~$\chi^{2}_{min}$~values have a decreasing trend 
with decreasing mass in the mass range of the 
black-hole. Fig.1~shows the photon flux versus ~$t_{*}$~graph for 
$M_{\mbox{x}}~\simeq~13M_{\odot}$, with the scale factor $b$  
corresponding to minimum ~$\chi^{2}$.

For this model
\be 
t_{*}= bt= 6~\nu R_{1}^{-2} t . \label{u20}~~
\ee 
With~$b\simeq~5.94~\times~10^{-7}~$ for $~M_{\mbox{x}}= 13~M_{\odot}$, 
Eq.(\ref{u20})~gives 
\be
\nu~R_{1}^{-2}~\simeq~8.9\times10^{-8}~\mbox{s}^{-1} \label{u22}~~.
\ee  

In the $\alpha$ disk model (Shakura~$\&$~Sunyaev~1973) the tangential 
stress can be characterized by the single parameter ~$\alpha$~ and the 
kinematic viscosity, $\nu$, may be written as 
\be 
\nu= -\alpha\frac{\upsilon_{\mbox{s}}^{2}}
{R\left(\frac{\d \Omega}{\d R}\right)} = 
\frac{2}{3} ~\alpha~\frac{\upsilon_{\mbox{s}}^{2}}{\Omega}~= 
\frac{2}{3} ~\alpha~\frac{kT}{m_{\mbox{p}}\Omega} \label{u28}
\ee

Now the question is when and where the viscosity parameter 
$~\alpha= (3/2)~\nu~(m_{\mbox{p}}\Omega/kT)$ becomes 
maximum in the disk. Since $\Omega\propto R^{-3/2}$~and~$kT\propto$ 
$R^{-3/4}$~ we may conclude that ~$\alpha$~ is always maximum at the inner 
disk. But ~kT~$\propto t_{*}^{-5/16}$ implies that the $~\alpha~$ 
parameter is increasing with time. We follow the luminosity evolution of 
the disk until the time when the inner disk has cooled down to a 
temperature characteristic of the quiescent state. The quiescent state 
temperatures, or temperature upper limits are in the range 
$\sim 0.1-0.3~$keV for most BSXTs (Tanaka $\&$ Shibazaki 1996). We shall 
use $0.3$~keV to derive constraints for $\alpha$. The time corresponding 
to this temperature is already longer than the time of the first mini 
outburst, after which the fit for the decay following the primary 
outburst is no longer relevant.  

Taking ~kT$\sim 0.3~$keV~ for the quiescent state inner disk 
temperature and $~\alpha\le~1~$~we obtain the limits 
\ba
\nu\le~1.8\times10^{11}~\mbox{cm}^{2}\mbox{s}^{-1} \nonumber
\ea
and using Eq.(\ref{u22})
\ba
~R_{1}\le 1.4\times 10^{9}~\mbox{cm}~\sim 95 R_{0}. \nonumber
\ea
$R_{1}~$ may be less 
than the outermost radius without effecting the fits as 
long as the X-ray photon flux 
coming from $~R>R_{1}$~is not significant compared to the $~R<R_{1}$~region. 
$R_{1}$ denotes the location of mass accumulation and release, 
and is not necessarily identical with the outermost radius. In other 
words we inject ~$\delta m$~ at $~R= R_{1}$~in our idealized model. 
We obtain $kT_{0}$(max)$\simeq 0.49~$keV from the fit and 
$c\simeq 1.15\times 10^{12}~$(cgs)  from Eq.(\ref{u16}). 
Using the relation 
$c= 4.73\times 10^{7}~(\nu R_{1}^{-2}~\delta m)^{1/4}~$(cgs)
obtained above for the uniform viscosity model, we have 
\be
\nu R_{1}^{-2}~\delta m\simeq~3.5\times10^{17}~~\mbox{(cgs)} \label{u23}~~.
\ee
Substituting the value of $\nu R_{1}^{-2}$~from Eq.(\ref{u22})~in
Eq.(\ref{u23})~we obtain the amount of mass injected initially at the 
outer disk
\be
\delta m\simeq~3.9\times 10^{24}~~\mbox{g} \label{u24}~.
\ee

The optically inferred mass transfer rates onto the outer disk of 
BSXTs are around $\sim 10^{15}-10^{16}~$g~s$^{-1}$~during the quiescent 
states. These values are similar to those obtained from 
the long term average,$~\langle\dot{M}_{\mbox{x}}\rangle$~, 
of the mass transfer in X-ray 
~outbursts  in order of magnitude (Tanaka~$\&$~ Shibazaki 1996, and 
references therein). This may 
show that most of the matter accreting onto the disk is accumulated at 
$R_{1}$ during quiescence and is used as a fuel for the next outburst. 
This is consistent with the observation that $\dot{M}_{x}$ during 
quiescence, which reflects the mass transfer reaching the inner disk, is 
much less than $\dot{M}_{\mbox{opt}}$. 
Only one outburst was detected for each of the sources GS 2000+25 and
GS/GRS 1124-68, so we use the average accretion rate 
$\langle\dot{M}_{\mbox{x}}\rangle\sim 2\times 10^{15}~$g~s$^{-1}$\ of 
another typical
BSXT, A0620-00, which has a recurrence time around 58 years (White et al 
1984), to estimate the recurrence times of our two sources. 
From  $~t_{\mbox{r}}\sim\delta m/\langle\dot{M}_{\mbox{x}}\rangle$~ 
the recurrence time is found to be around 62 years for GS~2000+25. 

On the other hand, using $~\alpha=(3/2)~\nu~(m_{\mbox{p}}\Omega/kT)$ 
with ~$kT=kT(R_{\mbox{out}})\simeq 
1.67\times 10^{-3}~$keV~where $R_{\mbox{out}}\sim 10^{3}\times R_{0}$~it is 
found that $~\alpha\simeq 5.7\times 10^{-3}$~ for the outer radius. 
At$~R=R_{1}$,~where mass accumulation occurs, $~\alpha$~ becomes 
$3.2\times 10^{-2}$. These 
values correspond to an inner disk temperature of around 0.3 keV. The 
inner disk temperature decreases to~0.3~keV at $t\sim$ 100~days. 

\begin{table*}
\caption[]{\label{t2}Results obtained from constant $~\nu~$model for GS
2000+25}
\begin{center}
\begin{tabular}{|lcc|cc|ccc|}
\hline
&&&\multicolumn{2}{c|}{constant $\nu$}&\multicolumn{2}{c}{$\nu=\nu
(h)$}&\\ \hline
~$M_{\mbox{x}}~~$~&~~~cos
i~~~&~~$R_{0}~~~$&~~~$kT_{0}$(max)~&~$\chi^{2}_{\mbox{min}}$
&~~~$kT_{0}$(max)~&~$\chi^{2}_{\mbox{min}}$& \\
$~(M_{\odot})$&&($10^{6}~$cm)&(keV)&&(keV)&& \\
\hline
6&0.26&5.3&0.87&23.5&0.87&19.2& \\
7&0.26&6.2&0.81&21.9&0.80&17.9& \\
8&0.47&7.1&0.66&17.8&0.65&14.6& \\
9&0.53&7.9&0.61&16.3&0.60&13.5& \\
10&0.58&8.8&0.56&15.1&0.57&12.7& \\  
11&0.61&9.7&0.53&14.2&0.54&12.0& \\
12&0.64&10.6&0.51&13.4&0.51&11.4& \\
13&0.66&11.5&0.49&12.8&0.49&10.9& \\
\hline
\end{tabular}
\end{center}
\end{table*}

\begin{figure*}
\vspace{7 cm}
\includegraphics{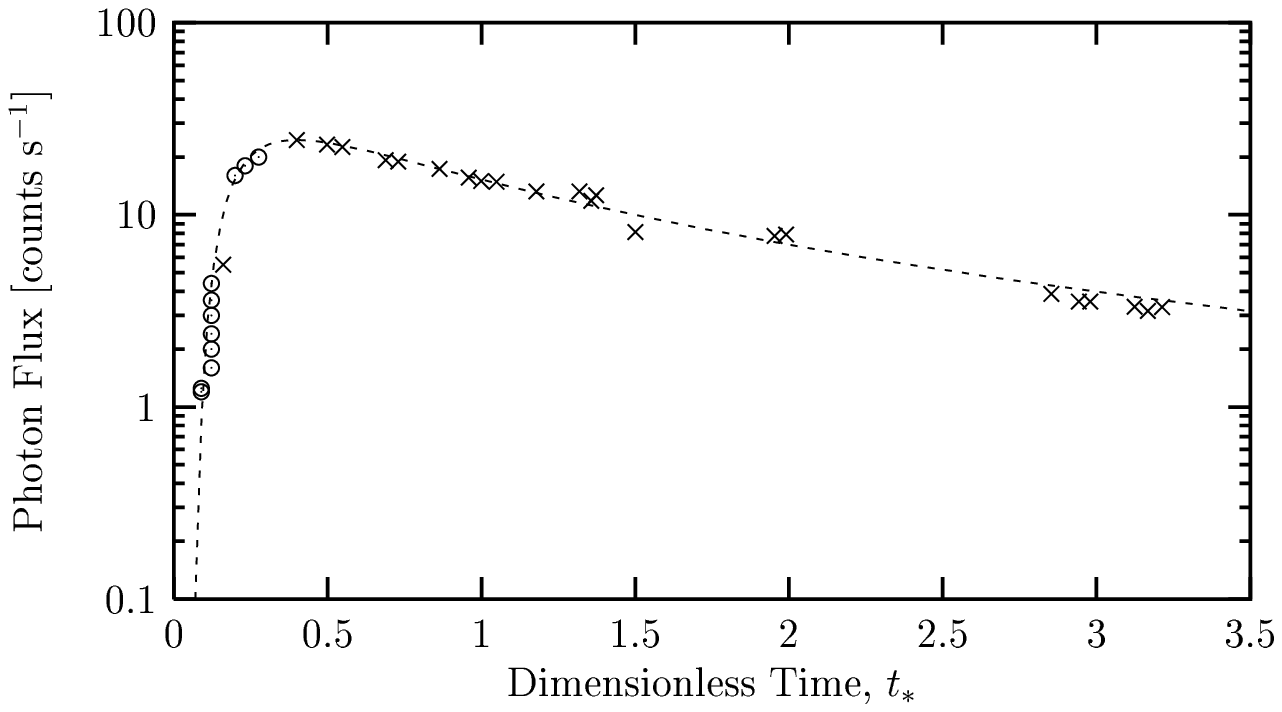}
\caption[]{Photon Flux versus dimensionless time for the  constant
$\nu$ model (for~GS~2000+25). The crosses are Ginga data points obtained
from S.Kitamoto. The quoted errors are small compared to the size of the
crosses. The circles are extra data points we scaled from Fig.3 in the
review paper by Tanaka \& Shibazaki (1996). These were not included in
fits or the $\chi^2$ calculations. They are shown here to
indicate the success of the model in encompassing the fast rise and
turnover phases of the light curve. The dashed line is the LP model fit.}

\vspace{8.5cm}
\end{figure*}

~~~

{\bf ii.~GS/GRS 1124-68}  

The distance of GS/GRS 1124-68 is around 3 kpc, the black-hole mass is 
$M_{\mbox{x}}$ $\sim 6 M_{\odot}$ (McClintock et al 1992). $kT_{0}$(max)$ = 
0.7~$keV produces the best fit to the 
Ginga observations of the outburst and decay light curve. For the~constant 
$\nu$ model, analysis similar to the one described above for 
GS~2000+25 gives the results 
\ba  
\nu R_{1}^{-2} \simeq 1 \times 10^{-7}~\mbox{s}^{-1} \nonumber  \\
\delta m \simeq 3 \times 10^{24}~\mbox{g} \nonumber \\
t_{\mbox{rec}} \simeq 49~\mbox{years} \nonumber \\
\nu \le 8.3 \times 10^{10}~\mbox{cm}^{2} \mbox{s}^{-1} \nonumber  \\
R_{1} \le 9.1 \times 10^{8}~\mbox{cm} \simeq 172~R_{0} 
\nonumber  \\ 
\alpha \le 5.85 (kT)^{-1} R^{-3/2}. \nonumber  
\ea
For GS/GRS 1124-68 we have taken  $M_{\mbox{x}}\sim 6 M_{\odot}$ 
(McClintock et al 1992) and from the 
best fit we have obtained $b\sim 6.13\times 10^{-7}$. 
When the inner disk temperature reduces to ~$\sim~0.3~$keV~the ~$\alpha~$ 
values are $~\sim 2 \times 10^{-2}~$ and $~\sim 5.3~\times~10^{-3}$~for 
$R=R_{1}~$and$~R=10^{3} R_{0}$~respectively. 
The best fitting model for the photon flux history is shown in Fig.2 
together with the data. The reduced 
~$\chi^{2}$~is found to be 123.7. Although this is a large value, we note 
that the magnitudes of the fluctuations of the observed 
photon flux data about a smooth decay curve are much higher 
than the errors  of the observed data points. As shown in Fig.2 the model 
reproduces the average rise and decay  behavior of the outburst light 
curve of GS/GRS 1124-68.

\begin{figure*}
\includegraphics{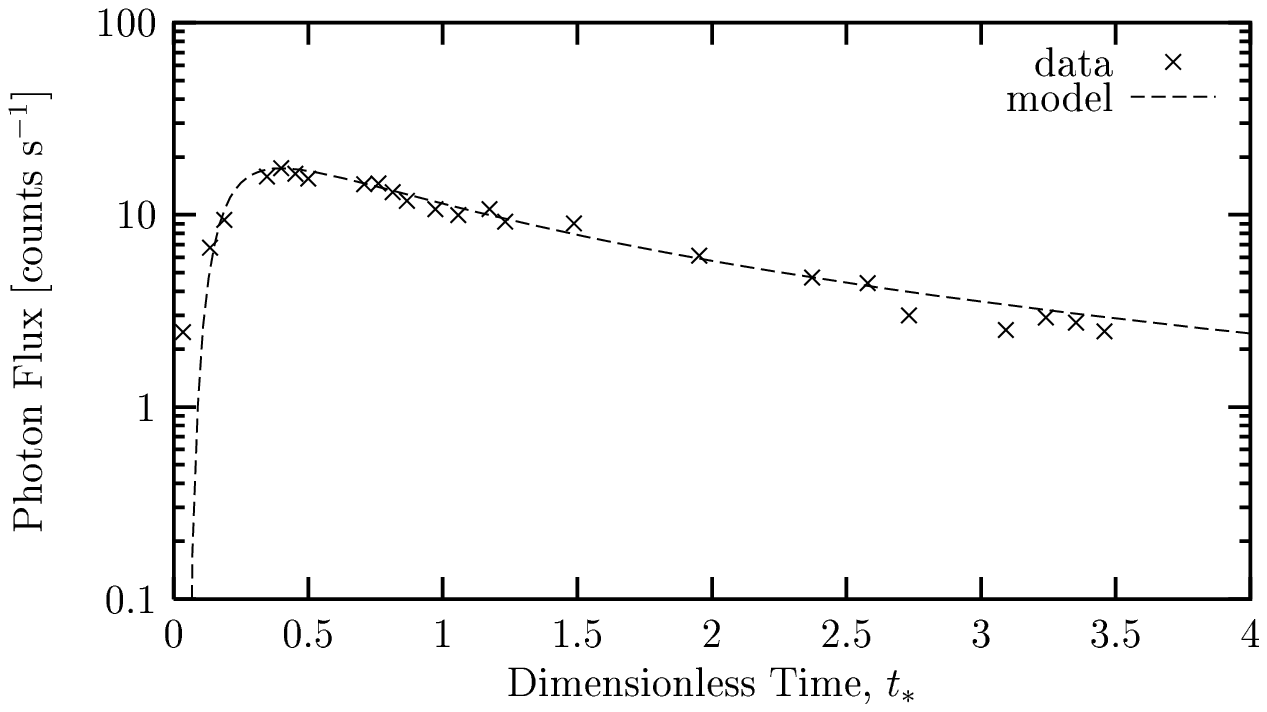}
\caption[]{Photon Flux versus dimensionless time
for the constant $\nu$ model (for~GS/GRS~1124-68).}
\end{figure*}

~~~~~

~~~~~~~~

{\bf MODEL II:~$\nu= \nu(h)$}

In this model, we allow $\nu$ to vary with radius (equivalently, with 
specific angular momentum, $h$) but not with temperature or time.

~~~

{\bf i.~GS 2000+25} 

We adopt the commonly used form of the viscosity parameter prescription  
$\alpha~=~\alpha_{0}~(z_{0}/R)^{n}$.~Cannizzo et al (1995)  
succeeded to reproduce the exponential type decays of the 
outbursts of BSXTs employing  this form of the viscosity parameter 
with $n=1.5$. The kinematic viscosity is 
$\nu= (2/3) (P/\rho) (\alpha /\Omega)$.
From the hydrostatic equilibrium in the z direction ~$P/\rho~\sim z_{0}^2$ 
$\Omega^{2}~$ and $z_{0}\sim\upsilon_{\mbox{s}}/\Omega~$, where 
$\upsilon_{\mbox{s}}^{2}= kT/m_{\mbox{p}}$~is the sound speed. Then
\ba
\nu= \frac{2}{3} \alpha_{0}\left(\frac{z_{0}}{R}\right)^{n} z_{0}^{2} \Omega 
\nonumber~~.
\ea
$\nu$ will be independent of temperature only for $n=-2$, which gives 
\be
\nu= \frac{2}{3} \alpha_{0}~(\Omega R^{2})= \frac{2}{3}\alpha_{0} h 
\label{u32}~~.
\ee

The specific angular momentum dependence of ~$\nu$~ is included in the 
general form of the solution for $~g~$ given by Eq.(\ref{u2}). For this 
choice of $~\nu~$ Eq.(\ref{u6}) becomes
\be
\frac{1}{2}\frac{(GM)^{2}}{h}~\alpha_{0}= 4l^{2}\kappa^{-2}~h^{2-(1/l)} 
\label{u33}~~. 
\ee
Then we obtain
\ba
l= 1/3 \nonumber
\ea
\be
\kappa^{-2}= \frac{9}{8}~\alpha_{0}~(GM)^{2} \label{u34}
\ee
and 
\be
t_{*}= \frac{9}{4}~\alpha_{0}~(GM)^{1/2}\frac{t}{R_{1}^{3/2}} \label{u35}~~.
\ee

Using Eqs.(\ref{u2}), (\ref{u9}), (\ref{u34})~$\&$~(\ref{u35})~ with 
~$l=1/3$~ we obtain the energy dissipation rate
\be
D\propto~t_{*}^{-4/3} R^{-3} 
exp\left[\frac{-[1+(R/R_{1})^{3/2}]}{2 t_{*}}\right] \label{u36}~~,
\ee
and  the temperature 
\ba
T&=& \left(\frac {D}{2\sigma}\right)^{1/4}~ \nonumber \\   
&=& c~t_{*}^{-1/3} R^{-3/4}~
exp\left[\frac{-[1+(R/R_{1})^{3/2}]}{8~t_{*}}\right] \label{u37}~~.
\ea
where
\be
c\simeq~1.1\times10^{11}~(\delta 
m~\alpha_{0}~R_{1}^{-3/2})^{1/4}~\mbox{(cgs)} \label{u38}~~. 
\ee

Now we follow the same procedure as we did for the constant ~$\nu$~ 
model. The temperature of the inner disk at the time corresponding to 
the maximum of the outburst is $~T_{0}\mbox{(max)}$~.
$(\d T_{0} / \d t_{*})= 0$ gives~$t_{*}\mbox{(max)}=0.375~$ 
and $T(R,t_{*})$~becomes
\ba
T(R,t_{*})\simeq T_{0}(max)~&R&_{0}^{3/4}~t_{*}^{-1/3}R^{-3/4}~\nonumber \\
&\times& exp\left[\frac{-[1+(R/R_{1})^{3/2}]}{8~t_{*}}\right] \label{u39}~~
\ea
where $~R_{0}= 3R_{g}= 3(2GM_{\mbox{x}}/c^{2})$.

\begin{table}
\caption[]{The results obtained from the first model}
\begin{center}
\begin{tabular}{|llll|}
\hline
~& GS 2000+25 & GS/GRS 1124-68 & \\
\hline
$R^{2} \nu^{-1}$ & $\simeq 1\times 10^{7}~$sec & $\simeq 1$
$\times10^{7}~$sec & \\
$\delta m$ & $\sim~3.9\times 10^{24}~$g & $\sim~3\times 10^{24}~$g & \\
$t_{\mbox{rec}}$ & $\simeq 62~$years &
$49~$years & \\ 
$\nu$(cm$^{2}$~s$^{-1}$) & $\le 1.8 \times 10^{11}$ &
$\le 8.3 \times 10^{10}$& \\
$R_{1}$ & $\le 1.4 \times 10^{9}~$cm &
$\le 9 \times 10^{8}~$cm & \\
& $\simeq 95 R_{0}$ & $\simeq 1.7 \times 10^{2} R_{0}$ & \\
$\alpha(R= R_{1_{max}})$ & $\le 3.2 \times 10^{-2}$ & $\le 2.0 \times
10^{-3}$ & \\
\hline
\end{tabular}
\end{center}
\end{table}

Once again we perform the numerical integration to 
construct the photon flux of the model disk as a function of time. By 
matching the maxima of the model and the observed light curve, we obtain 
different $kT_{0}$(max) values for 
different masses in the range $~6<M_{\mbox{x}}/M_{\odot}<13.9~$. Later using 
$~t_{*}$= $bT~$ we obtain $~b~$ values producing minimum $~\chi^{2}~$for 
each mass tested. The results are in Table(2). From the best fit we 
obtain $~M_{\mbox{x}}=13M\odot$, $~R_{0}=1.15\times10^{7}~$cm, 
$kT_{0}\mbox{(max)}= 0.49$ keV, and  $b\simeq 4.63\times 10^{-7}$. From 
Eq.(\ref{u35})  
\be
b\simeq~9.34\times10^{13}~\alpha_{0}~R_{1}^{-3/2} \label{u40}
\ee
we obtain
\be
\alpha_{0}~R_{1}^{-3/2}= 7.29\times10^{-21}~cm^{-3/2}. \label{u41}
\ee
The corresponding ~$T_{0}$(max)~ produces ~$c\simeq 1.17\times 
10^{12}~$and, using Eqs.(\ref{u38}), and (\ref{u41})~we find
\ba
\delta m~\alpha_{0}R_{1}^{-3/2}\simeq 1.6\times 10^{4} \mbox{g~cm}^{-3/2} 
\label{u42} \\
\delta m\simeq~3.3\times10^{24}~\mbox{g} \nonumber.
\ea
Dividing ~$\delta m~ $by$ ~\dot{M}_{\mbox{opt}}$
$\sim~ \langle\dot{M}\rangle\simeq$~ 
$2\times 10^{15}~$g~s$^{-1}~$ we obtain the recurrence time
\ba
t_{r}\sim~52~\mbox{years} \nonumber
\ea
The viscosity parameter $\alpha$, with $n=-2$ and 
$~z_{0}=\upsilon_{\mbox{s}}/\Omega$~ is 
\be
\alpha= \alpha_{0}\frac{R^{2}\Omega^{2}}{\upsilon_{\mbox{s}}^{2}} 
\label{u44} \ee
with ~$\upsilon_{\mbox{s}}^{2}= kT/m_{p}~$ and 
$~\Omega= (G M_{\mbox{x}})^{1/2} R^{-3/2}$ 
\be
\alpha= \alpha_{0}\frac{GM_{\mbox{x}}m_{p}}{kT~R}. \label{u45}
\ee
Since $~kT\propto R^{-3/4}~$ 
the viscosity parameter $\alpha\propto R^{-1/4}$ ~becomes 
maximum at the inner disk. 
Then as we have found in the previous case ~$\alpha$~ increases with time 
until $~kT_{0}\sim0.3~$keV. The inner disk temperature reduces to
$kT_{0}\sim0.3~$keV~in about 100 days. Eq.(\ref{u44})~with $~\alpha=$ 
$\alpha_{max}\simeq 1~$ ,~$kT\sim~0.3~$keV~and 
$~R= R_{0}= 1.15\times 10^{7}~$cm gives $\alpha_{0}\le 1.9\times 10^{-6}$. 
Using Eq.(\ref{u41}) we find
\ba
R_{1}\le~5.3\times 10^{9}~\mbox{cm}~\simeq~461 R_{0}  \nonumber~.
\ea
As stated before we may consider ~$R_{1}$~to be the radius where the mass 
accumulates. This ~$R_{1}$~value is nearly 5 times greater 
than the one obtained in the first model. 

When the inner disk temperature is around ~0.3 keV~using 
~Eq.(\ref{u39})~ we 
find that the temperature is $\sim 1.7\times 10^{-3}~$keV~at$~R\sim 10^{3}$ 
$R_{0}$~and ~$\sim3.0\times10^{-3}$~keV~ at $~R=R_{1}\sim 461~R_{0}$.  
The corresponding $~\alpha~$ values are ~0.18~and~0.21~respectively, which 
are plausible values.

\begin{figure*}
\vspace{8.5 cm}
\includegraphics{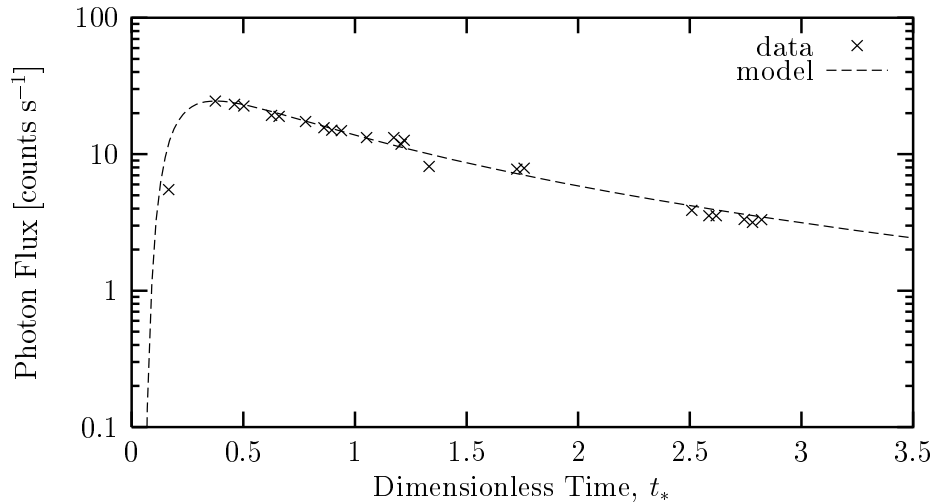}
\caption{Photon flux versus dimensionless time for the
$\nu=\nu(h)$ model (for~GS~2000+25).}
\vspace{1.0cm}
\end{figure*}

~~~

{\bf ii.~GS/GRS 1124-68}

For this model $kT_{0}(max)= 0.7~$keV~ produces the best fit to the 
observed data. Following the steps described above for GS~2000+25 
we have obtained a similar fit with the reduced 
~$\chi^{2}=105.3~$(Fig.4). The results are:
\ba
\alpha_{0} \times R_{1}^{-3/2} \simeq 8.8 \times 10^{-21}~\mbox{cm}^{-3/2} 
\nonumber  \\ 
\delta m \simeq 2.37 \times 10^{24}~\mbox{g} \nonumber \\
t_{\mbox{rec}} \simeq 38~\mbox{years} \nonumber \\
\alpha_{0} \le 1.9 \times 10^{-6} \nonumber \\
R_{1} \le 3.63 \times 10^{9}~\mbox{cm} \simeq 685~R_{0} 
\nonumber \\ 
\alpha \le 2.5 \times 10^{-3} (kT R)^{-1} \nonumber 
\ea
Since $~M_{\mbox{x}}/R_{0}=c^{2}/6\mbox{g}$ is constant 
the upper limit for $\alpha_{0}$ is 
the same as that of GS~2000+25 (Eq.\ref{u45}). 
The recurrence time is similar to that of the first model. 
When the inner disk temperature reduces to ~$\simeq 0.3~$keV~the 
$~\alpha~$values are $~\simeq 0.19 ~$ and 
$~\simeq~0.17~$for ~$R=R_{1}~$and 
$~R= 10^{3} \times R_{0}~$ respectively. 

From the Roche-lobe geometry the distance from the center of
the compact object to the inner Lagrangian point $L_{1}$,
$R_{L_{1}}\sim 2\times 10^{11}~$cm for both sources. 
When we set $R_{1}\sim R_{L_{1}}$ we find the 
$\alpha_{\mbox{max}}\ll 1$ which is not plausible. 
We may conclude that the radius $R_{1}$ at which 
the mass accumulation occurs is 
around or less than $2\times 10^{-2}$ times the disk's outer radius, 
if the outer radius $R_{\mbox{outer}}$ is 
of the same order as $R_{L_{1}}$. 
If we take $R_{1}~\sim~R_{\mbox{outer}}$, our inferred values of $R_{1}$ 
would mean that $R_{L_{1}}$ is at least fifty times greater than 
$R_{\mbox{outer}}$.

\begin{table}
\caption{The results obtained from the second model}
\begin{center}
\begin{tabular}{|llll|}
\hline
~& GS 2000+25 & GS/GRS 1124-68 & \\
\hline
$\delta m$ & $\sim~3.3\times 10^{24}~$g & $\sim~2.4\times 10^{24}~$g & \\
$t_{\mbox{rec}}$ & $\sim 52~$years & $\sim 38~$years & \\
$\alpha_{0}$ & $\le 1.9 \times 10^{-6}$ &
$\le 1.9 \times 10^{-6}$ & \\
$R_{1}$ & $\le 5.3 \times 10^{9}~$cm &
$\le 3.6 \times 10^{9}~$cm & \\
& $\simeq 461 R_{0}$ & $\simeq 685 R_{0}$ & \\
$\alpha(R= R_{1_{max}})$ & $\le 0.21$ & $\le 0.19$ & \\
$\alpha(R= 10^3~ R_{0})$ & $\le 0.18$ & $\le 0.17$ & \\
\hline
\end{tabular}
\end{center}
\end{table}

\begin{figure*}
\vspace{7.5cm} 
\includegraphics{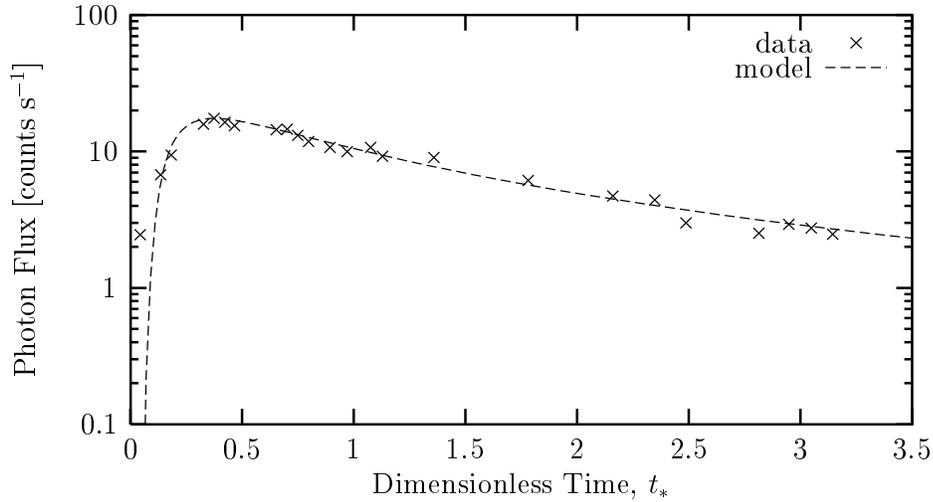}
\caption[]{Photon flux versus dimensionless time for the
$\nu = \nu(h)$ model (for~GS/GRS~1124-68).}
\vspace{7.0cm}
\end{figure*}

\begin{figure*}
\vspace{2.0cm}
\includegraphics{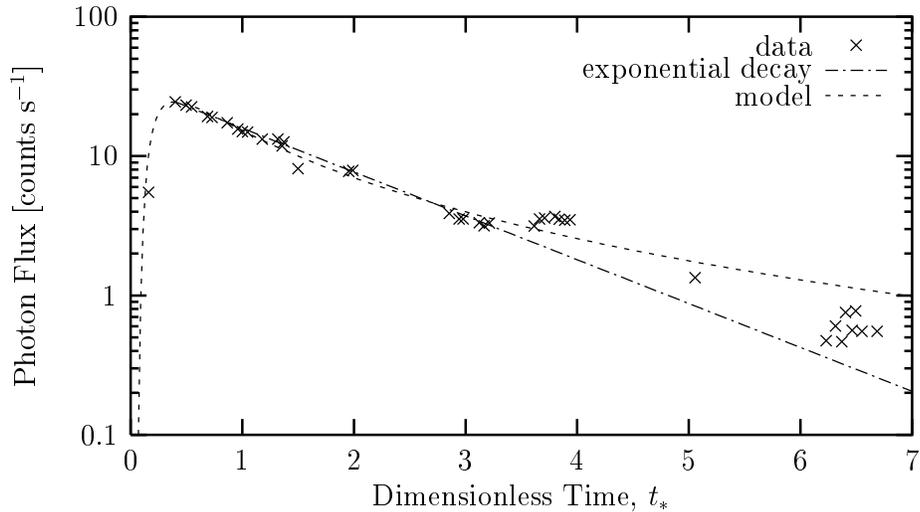}
\caption[]{Photon flux versus dimensionless time for the constant
$\nu$ model for~GS~2000+25 (dashed line) together with an
exponential decay curve (dot-dashed line) with a time constant 30
days. The reduced $\chi^{2}_{min}$ values are 5 and 12.8 for the
exponential decay and the model respectively, for fits up to the
onset of the first mini outburst ($t_{*}<3.5$). It is seen that
exponential decay models for the mini outburst decays will superpose
on the continuing decay of the main outburst. By contrast, LP fits
to the mini outbursts require that each mini outburst "zeroes" the
continuing decay of the main outburst, and starts with the same
initial conditions as the main outburst.}
\vspace{0.5 cm}
\end{figure*}

\section{Discussion and Conclusions}

We have addressed the general characteristics of the outburst light 
curves of black-hole soft X-ray transients (BSXTs), through a study of 
the typical BSXTs GS 2000+25 and GS/GRS 1124-68.  
The Green's function solution of the accretion disk dynamics model of 
Lynden-Bell $\&$ Pringle (1974) (LP), with a delta function 
initial mass distribution, produces a rise and decay pattern similar to 
those of BSXTs. It is most remarkable that this simple dynamical model 
gives both the rise and the subsequent decay. We have employed this solution 
adopting a 
delta function initial mass distribution, a mass $\delta m$ deposited 
in a ring at radius $R_{1}$ in the disk, for two different time 
independent viscosity laws. The LP solution restricts the viscosity 
to be a time independent power law function of the radial distance 
from the center of the disk.

For the first model we chose a constant kinematic viscosity $\nu$ for 
simplicity. For the second model we adopted 
the commonly used viscosity parameter 
prescription $\alpha~=~\alpha_{0}~(z_{0}/R)^{n}$~ and chose ~$n=-2$~ to 
remove the time varying temperature dependence of ~$\nu$. This leads to 
~$\nu=\nu(h)$, where h is the specific angular momentum. 

For both models $kT_{0}$(max), the inner disk temperature at the moment 
that the light curve for photon count rate reaches its maximum, is the free 
parameter 
together with $cos~i$. We adjusted $kT_{0}$(max) to match the maxima of 
the model and the observed photon 
flux light curves. Next, from $t_{*}=b~t$ where $
t_{*}$ is dimensionless 
time and $t$ is real time, the best fit to the 
observed photon flux was obtained by adjusting the constant $b$. 
For GS~2000+25 this procedure was performed 
tracing the possible mass range, 
$6 M_{\odot} < M_{\mbox{x}} < 13.9 M_{\odot}$, corresponding to the possible 
inclination angle range, $~47^{\circ}<~i~<75^{\circ}$ (Harlaftis at al 1996). 
For GS/GRS 1124-68 the black-hole mass is $M_{\mbox{x}}\sim 6 M_{\odot}$ 
(McClintock et al 1992).          
For this source we tried four different $cos~i$ values (0.25, 0.5, 0.75, 1.0)
keeping $M_{\mbox{x}}$ constant and followed the same steps to obtain the 
best fit. The results are in Tables (2$-$4)

For both sources the fits are applied to the data until the onset of 
the first mini outburst. The LP model fits the secular decay in the light 
curve. The large $\chi^{2}$ values reflect the excursions about the mean 
secular decay, which the model does not address. Figs. (1$-$4) show that 
both models reproduce the average characteristics of the outburst light 
curves quite well. The requirement of the model that the kinematic 
viscosity is independent of time and therefore also of the temperature 
leads to an $\alpha$ parameter which increases with decreasing 
temperature. The usual disk instability models require the opposite 
trend of $\alpha$ increasing with increasing temperature. Our time 
independent viscosity models are artificial and may be considered as an 
effective representation of real viscosity behavior during the hot 
states of the accretion disks of BSXTs. Empirically the LP models can 
explain the rise and the decay following the outburst. 
Our approach here shows that once sudden 
mass release is triggered at some $R_{1}$, for example by a disk
instability, the subsequent light curve can be understood as the
dynamical evolution of the disk with viscosity.
It seems plausible that the viscosity
mechanism must have a temperature dependence at least to start the outbursts.
The success of the model means that the real disk dynamics is insensitive 
to the temperature dependence of the viscosity, or that a constant 
effective viscosity LP model is a close approximation  to the real 
luminosity evolution with the integrated effects of variable and 
temperature dependent viscosity. To investigate the relation between the 
LP models and disks with variable viscosity is beyond the scope of the 
present paper, and is to be considered in subsequent work. 
 
The mass dumped in the outburst is $\delta m \sim 10^{24}$~g in all our 
model fits, for both sources. This is consistent with accumulation at 
$\langle \dot{M}_{\mbox{x}}\rangle \sim$ 
$\langle\dot{M}_{\mbox{x}}\rangle_{0620}$~ for recurrence times 
$t_{r}\sim \delta m/\langle\dot{M}_{\mbox{x}}\rangle\sim 50$~yrs. We expect 
that the mass release is triggered by a disk instability. The large 
amounts of mass release $\delta m$ imply surface densities 
$\Sigma~\sim~(\delta m/R_{1}^{2})\sim 10^{4}-10^{5}~$ g~cm$^{-2}$ which 
are far in excess of the critical (maximum) surface density 
values for disk instability models, 
\ba
\Sigma_{max}=11.4~ \mbox{g~cm}^{-2}~R_{10}^{1.05}  
\L( \frac{M}{M_{\odot}} \R)^{-0.35} \alpha^{-0.86}   \nonumber
\ea 
(Shafter at al 1986). The mass accumulation without triggering an 
instability then requires small $\alpha$ values $\sim 10^{-4}-10^{-5}$ 
in the quiescent disk. At such $\alpha$ the quiescent disk could be 
optically thin.    

There is no complete picture yet which explains all properties of BSXTs,
in particular the triggering mechanism for the outbursts which drives the
accretion disk from the quiescent state to the outburst and back to the
quiescent state. In a recent study by Cannizzo et al~(1995)~
exponential decay patterns are produced by the disk limit cycle mechanism,
but sudden rise of the outbursts and their long recurrence times are not
addressed.

The second (and in two sources, third) outbursts have 
rise and decay patterns which can be fitted with exponentials 
with time constants similar to those of the main outburst. 
In addition to this similarity they also pose the problem of 
understanding the burst repetition time (Augusteijn et al 1993; 
Chen et al 1993; Mineshige 1994). 
What is the source of the mass producing the second  outburst? The 
data following the second mini outburst does not allow a detailed fit 
with the LP model. Assuming that the amplitude of the burst is 
proportional to the accumulated mass, $\delta m$, and scaling with the 
primary outburst we find $\delta m$ values are $\sim 3\times 10^{23}~$g 
and $\sim 2.5\times 10^{23}~$g for the second 
mini outbursts in GS/GRS~1124-68 and GS~2000+25 respectively. Using the 
time interval between the main and the second mini outburst $\sim 80~$
days, we find the extra $\dot{M}$ to supply the $\delta m$ of the 
second mini outburst is $\sim 4.5\times 10^{16} $g~s$^{-1}$ for 
GS/GRS~1124-68 and $\sim 3.5\times 10^{16}~$g~s$^{-1}$ for GS~2000+25. This 
is nearly an order of magnitude greater than 
the long term average accretion rate, $\langle\dot{M}\rangle$. 
Two possibilities may be considered: (1) All the 
accumulated mass is not released during the main outburst. The mass 
released in the second outburst is "leftover mass" and not accumulated 
between the main outburst and the second outburst.  (2) The mass 
flow from the companion is enhanced by the X-rays coming from the inner 
disk in the outburst; or both possibilities run together. The X-ray 
heating of the disk may be important to trigger the mini outbursts 
whatever the source of the mass creating the second and the third 
instabilities. 

The time between the main and mini outbursts ($\sim$ few months)must be a
characteristic time scale of the "trigger". 
The response time scale of the upper atmosphere of the secondary is too
short ( $<\atop{\sim}$ $10^{3}$ s)~ (Chen et al 1993 (CLG)) while  the 
viscous time scale is of the order of a week. 
The shielding of the region around the inner 
Lagrangian point, $L_{1}$, has been proposed to explain this time 
interval (CLG). If the geometrical shielding 
should end gradually, the abrupt rise of the first mini outburst 
at a particular time remains unexplained. Two different approaches were 
proposed by Mineshige (1994). An optically thick Compton 
cloud above and below the disk which becomes optically thin in a very short 
time scale just before the first mini outburst may explain the 
first mini outburst by the response of the companion to the X-rays coming 
from the inner disk. Alternatively a second thermal instability 
would be triggered at the outer disk when the strong  
X-ray heating of the disk has raised the temperature and 
decreased the critical density for the 
trigger sufficiently (transient recession of the cooling front). In the 
former scenario 
one still has to explain the secondary mini outburst by invoking a 
different mechanism. 
and also exhibits similar rise and decay time scales. 
Similar time interval and characteristics of all the 
outbursts seem to imply a unique mechanism responsible for the triggering 
of the outbursts. 

It is well known that simple exponential decays after the 
outburst maximum describe the data well. The mini outburst decays can 
also be fitted with exponential decays with the same time constants as 
required by the main outburst(Fig.5). However, the fast rise and the 
exponential 
decays (FRED) are put together as separate pieces of the fit rather than 
being part of a single dynamical model as in the present model. Our LP 
models fit the main outburst decay quite well. There is an interesting 
difference from FRED models when it comes to trying to fit the data 
incorporating the mini outbursts and their decays. In FRED models, the 
data stream can be fitted well by a superposition of FRED models for the 
main outburst and the mini outbursts. For our LP model fits, by contrast, 
when the fit to the main outburst data is extended past the mini 
outburst(s), it is seen that the model gives a count rate that is greater 
than the observed count rates, the deviation starting from the onset of 
the mini outburst. If it is assumed that the ongoing relaxation in the 
disk after the main outburst is stopped, and the onset of the mini 
outburst involves an instability that resets the disk to conditions 
similar to sudden mass release from a local accumulation, then the LP 
model can be applied to the mini outburst(s) and their decays.       
 
Augusteijn et al (1993) drew attention to the similarity of 
the main outburst decay, and the mini outburst decays. They related this 
to a feed-back model invoking modulation of the mass transfer from the 
companion by irradiation from the outburst. The present approach 
identifies the similarity with repeated conditions of mass accumulation 
and release in the disk itself. The trigger of the mass release could 
be a disk instability. The similarity of main outburst and mini 
outbursts, requires small scale repetitions of the main outburst with 
similar initial conditions, rather than superposition or convolution with 
the disk state as evolved from the main outburst's 
decay. The presently available data do not allow a detailed fit of 
the mini outbursts with the LP models. 

The behavior of  ~GS~2000+25~ and ~GS/GRS~1124-68~ are similar to those of 
A0620-00,~and GRO J0422+32. The conclusions may be extended to them as 
well. These ideas will be developed and a similar detailed study for 
the other sources will also be attempted in future work.

~~~~

{\it \bf Acknowledgements}
This work started from discussions with our late colleague Jacob Shaham.
We thank S.Kitamoto for providing GINGA outburst data for the sources 
GS~2000+25 and GS/GRS~1124-68, and the referee S.Mineshige for his 
helpful comments. We thank the Scientific and Technical Research
Council of Turkey, T\"UB\.ITAK, for support through the grant TBAG 
\"U-18. \"U.Ertan thanks T\"UB\.ITAK for a doctoral scolarship. 
M.A.Alpar acknowledges support from the Turkish Academy of Sciences.

\appendix
\section*{Appendix}

Here we summarize, following Lynden-Bell $\&$ Pringle 1974 (LP), 
the general arguments of the angular
momentum and the energy flow in the thin accretion disks. Consider a 
thin annular section of a disk of width dR at radius R from 
the center. The torque or couple acting on the inner edge of the annulus is
\bc
$$g(R)= 2\pi R^{2}~\nu~\Sigma~2~A(R) . \eqno(A.1) $$ 
\ec
where A(R) is the local rate of shearing  
\bc
$$A= -\frac{1}{2}~R~\frac{d\Omega}{dR}. \eqno(A.2) $$  
\ec
$\Omega$~ is the angular velocity about the center, $\nu(R)$ is the    
kinematic viscosity and ~$\Sigma(R)$~ is the surface density. Viscous  
forces are proportional to velocity gradients, and viscous
torques are proportional to angular velocity gradients (shear).

We may write the equation of motion of the annulus at radius R in   
the form
\bc
$$\frac{D}{Dt}~(h~dm)= -\frac{\partial g}{\partial R}~dR. \eqno(A.3)$$
\ec
where ~$h= \Omega R^{2}$, ~$dm= \Sigma~2\pi R~dR$ ~and~
$(D/Dt)= (\partial/\partial t)+{\bf U}.\nabla~$ is the total
(convective or Lagrangian) time derivative following a fluid element. 
Eq.($A.3$) states that rate of change of angular momentum is
given by the torque.

Since $D(dm)/Dt=0$
\bc
$$2\pi R \Sigma \frac{Dh}{Dt}= -\frac{\partial g}{\partial R}=
\frac{\partial}{\partial R} \left(2\pi R^{3} \Sigma\nu 
\frac{\partial \Omega}{\partial R}\right). \eqno(A.4) $$
\ec
If the specific angular momentum $h= \Omega R^{2}$ is independent of time
Eq.($A.4$) becomes
\bc
$$F \frac{dh}{dR}= -\frac{\partial g}{\partial R}= 
\frac{\partial}{\partial R}
\left(4\pi R^{3}\Sigma\nu ~\frac{\partial\Omega}{\partial R}\right)
\eqno(A.5)$$
\ec
where $F$ is the outward flux of matter through the radius R. For a
Keplerean disk we can take
\bc
$$\Omega= (GM)^{1/2}~R^{-3/2} \eqno(A.6)$$
\ec
where $M$ is the mass of the star at the center of the disk.  

Energy generated by the matter flowing into
the gravitational well is redistributed or convected by
$g\Omega$, and the remainder is dissipated by
\bc
$$D= \frac{1}{2\pi R}~g~ \L(-\frac{d\Omega}{dR}\R). \eqno(A.7)$$
\ec

The continuity equation for the fluid is
\bc
$$\frac{\partial\S}{\d t}~+~\frac{1}{2\pi R}~\frac{\d F}{\d R}= 0
\eqno(A.8)$$
\ec
where $~\Sigma~$ is the surface density of the disk.
Using Eqs.($A.4 \& A.8$)   
\bc
$$\frac{\partial^{2} g}{\partial h^{2}}= \frac{\d}{\d t}
~\left[\frac{g}{2A\nu R~(dh/dR)}~\right] \eqno(A.9)$$
\ec
is obtained. Lynden-Bell $\&$ Pringle (1974)~express the denominator as
\bc
$$2A\nu R~\frac{dh}{dR}= 4l^{2}~\kappa^{-2}~h^{2-(1/l)} \eqno(A.10)$$
\ec
where $\kappa$~and~l~ are constants. Eq.($A.10$) is valid as long
as $\nu$ is constant or varies as a function of R, and can be used to
parametrize the $\nu$ and h laws. Eq.($A.9$) reduces to
\bc
$$\frac{\partial^{2} g}{\partial h^{2}}=
\frac{1}{4}~\left(\frac{\kappa}{l}\right)^{2} h^{(1/l)-2}~\frac{\d g}{\d t}. 
\eqno(A.11)$$
\ec
Resolving Eq.($A.11$) into modes where $g~\propto~e^{-st}$ and
setting $k^{2}= \kappa^{2} s$
\bc
$$\frac{\d^{2} g}{\d h^{2}}~+~\frac{1}{4}\left(\frac{k^{2}}{l^{2}}\right)~
h^{(1/l)-2} g= 0 \eqno(A.12)$$
\ec
and changing the variables $x=~h^{1/2l}$ and $g_{1}=~x^{-l} g$ in
Eq.($A.12$) gives a transformation of Bessel's equation 
\ba
\frac{\d^{2} g_{1}}{\d x^{2}}~+x^{-1}\frac{\d g_{1}}{\d x}~
+~\left(k^{2}-\frac{l^{2}}{x^{2}}\right)g_{1}= 0 \nonumber
\ea
with solution given by
\bc
$$g= e^{-st}~(kx)^{l}~[A(k)J_{l}(kx)~+~B(k)J_{-l}(kx)]. 
\eqno(A.13)$$
\ec

The general solution of Eq.(A.11) is given by a convolution of
such modes. Then
\bc
$$g= \int_{0}^{\infty}~exp\left(-\frac{k^{2}
t}{\kappa^{2}}\right)~(kx)^{l}~[A(k)J_{l}~+~B(k)J_{-l}]~dk \eqno(A.14)$$
\ec
where B(k)=0 for the central couple to be zero, ~$g(x=0)=0.~A(k)$~ is given
by the inverse transform
\bc
$$A(k)= \int_{0}^{\infty}~g(h,0)~J_{l}(kx)~(kx)^{1-l} dx. \eqno(A.15)$$
\ec

When there is no central couple as in the case of BSXTs any solution may
be considered as made up of elementary solutions whose initial density
distributions are of the form
\bc
$$\Sigma(R_{1},R,0)= (2\pi R_{1})^{-1}~\delta(R-R_{1}) \eqno(A.16)$$
\ec
and the corresponding couple is
\bc
$$g(h_{1},h,0)= 2l~\kappa^{-2}~x_{1}^{2l-1}~\delta(x-x_{1}). \eqno(A.17)$$
\ec
For the special initial condition in Eq.($A.17$), using Eq.($A.15$)
\ba
A(k)= 2l\kappa^{-2}~x_{1}^{l}~k^{1-l}~J_{l}(kx_{1}). \nonumber
\ea
Substituting $A(k)$ in Eq.($A.14$) with B(k)=0
\bc
$$g= 2l\kappa^{-2}~\frac{(x_{1} x)^{l}}{x_{1}^{2}
t_{*}}~exp-\left[\frac{(x_{1}-x)^{2}}{2t_{*}
x_{1}^2}\right]~F_{l}\left(\frac{x/x_{1}}{t_{*}}\right) \eqno(A.18)$$ 
\ec
where $t_{*}= 2\kappa^{-2} t/x_{1}^2,$~ $~F_{l}(z)= e^{-z}~I_{l}(z)$ and
$I_{l}(z)$ is the Bessel function of imaginary argument.
   
For the general initial conditions $g(h_{1},0)$ the general solution is
given by

$$g(h,t)= \int_{0}^{\infty}~g(x_{1}^{2l},
0)\kappa^{-2}~\frac{x_{1}^{-(1+l)}~x^{l}}{t_{*}}~
exp-\left[\frac{(x_{1}-x)^{2}}{2t_{*}x_{1}^2}\right]~$$\\
$$\times F_{l}\left(\frac{x/x_{1}}{t_{*}}\right)~dx_{1}$$ \\
$$\eqno(A.19)$$

(Lynden-Bell $\&$ Pringle, 1974)(LP). Substituting  this solution in 
Eq.($A.7$) and integrating throughout the disk one can obtain the 
total luminosity of the disk for the general initial conditions 
$g(h_{1},0)$. Instead we use the solution given by Eq.($A.18$) 
which represents a delta function initial mass distribution at a radial 
distance $R=R_{1}$, which may be considered as the outer radius of the 
disk. The luminosity curve obtained by LP with a delta function initial 
mass distribution is very similar to those of BSXTs. 

\end{document}